\documentclass[aps, reprint, prl, superscriptaddress]{revtex4-1}

\usepackage{hyperref}
\usepackage{graphicx}
\usepackage{amsmath}
\usepackage{wasysym}
\usepackage{amsfonts}
\usepackage{color}
\usepackage{verbatim}
\usepackage{upgreek}

\usepackage[final]{changes}

\newcommand{\beq}{\begin{equation}}
\newcommand{\eeq}{\end{equation}}
\newcommand{\bea}{\begin{eqnarray}}
\newcommand{\eea}{\end{eqnarray}}
\newcommand{\bal}{\begin{align}}
\newcommand{\eal}{\end{align}}

\newcommand{\WSe}{WSe$_2$}
\newcommand{\Si}{$\mathrm{\Sigma}$}
\newcommand{\Ga}{$\mathrm{\Gamma}$}

\begin{document}

\title{Generation and evolution of spin-, valley- and layer-polarized excited carriers in inversion-symmetric \WSe}

\author{R. Bertoni}
\affiliation{Fritz-Haber-Institut der Max-Planck-Gesellschaft, Faradayweg 4-6, 14195 Berlin, Germany}
\author{C. W. Nicholson} 
\affiliation{Fritz-Haber-Institut der Max-Planck-Gesellschaft, Faradayweg 4-6, 14195 Berlin, Germany}
\author{L. Waldecker} 
\affiliation{Fritz-Haber-Institut der Max-Planck-Gesellschaft, Faradayweg 4-6, 14195 Berlin, Germany}
\author{H. H\"ubener}
\affiliation{Nano-Bio Spectroscopy Group and ETSF, Universidad del Pais Vasco, CFM CSIC-UPV/EHU, 20018 San Sebastian, Spain}
\author{C. Monney}
\affiliation{University of Zurich, Department of Physics, Winterthurerstrasse 190, 8057 Z{\"u}rich, Switzerland  }
\author{U. De Giovannini}
\affiliation{Nano-Bio Spectroscopy Group and ETSF, Universidad del Pais Vasco, CFM CSIC-UPV/EHU, 20018 San Sebastian, Spain}
\affiliation{Dipartimento di Fisica e Chimica, Universit\`a degli Studi di Palermo, Via Archirafi 36, I-90123 Palermo, Italy
}
\author{M. Puppin} 
\affiliation{Fritz-Haber-Institut der Max-Planck-Gesellschaft, Faradayweg 4-6, 14195 Berlin, Germany}
\author{M.~Hoesch} 
\affiliation{Diamond Light Source, Harwell Campus, Didcot OX11 0DE, United Kingdom}
\author{E. Springate} 
\affiliation{Central Laser Facility, STFC Rutherford Appleton Laboratory, Harwell Campus, Didcot OX11 0QX, United Kingdom}
\author{R. T. Chapman} 
\affiliation{Central Laser Facility, STFC Rutherford Appleton Laboratory, Harwell Campus, Didcot OX11 0QX, United Kingdom}
\author{C. Cacho} 
\affiliation{Central Laser Facility, STFC Rutherford Appleton Laboratory, Harwell Campus, Didcot OX11 0QX, United Kingdom}
\author{M. Wolf} 
\affiliation{Fritz-Haber-Institut der Max-Planck-Gesellschaft, Faradayweg 4-6, 14195 Berlin, Germany}
\author{A. Rubio}
\affiliation{Nano-Bio Spectroscopy Group and ETSF, Universidad del Pais Vasco, CFM CSIC-UPV/EHU, 20018 San Sebastian, Spain}
\affiliation{Max Planck Institute for the Structure and Dynamics of Matter and Center for Free-Electron Laser Science, Notkestra{\ss}e 85, 22761 Hamburg, Germany}
\author{R. Ernstorfer}
\email{ernstorfer@fhi-berlin.mpg.de}
\affiliation{Fritz-Haber-Institut der Max-Planck-Gesellschaft, Faradayweg 4-6, 14195 Berlin, Germany}

\begin{abstract}

We report the spin-selective optical excitation of carriers in inversion-symmetric bulk samples of the transition metal dichalcogenide (TMDC) \WSe.  Employing time- and angle-resolved photoelectron spectroscopy (trARPES) and complementary time-dependent density functional theory (TDDFT), we observe spin-, valley- and layer-polarized excited state populations upon excitation with circularly polarized pump pulses, followed by ultrafast ($<100$ fs) scattering of carriers towards the global minimum of the conduction band. TDDFT reveals the character of the conduction band, into which electrons are initially excited, to be two-dimensional and localized within individual layers, whereas at the minimum of the conduction band, states have a three-dimensional character, facilitating inter-layer charge transfer. These results establish the optical control of coupled spin-, valley- and layer-polarized states in centrosymmetric materials with locally broken symmetries and suggest the suitability of TMDC multilayer and heterostructure materials for valleytronic and spintronic device concepts.

\end{abstract}

\maketitle
\date{\today}

Manipulation of spin and valley degrees of freedom is a key step towards realizing novel quantum technologies \cite{Mak2012,Zeng2012,Gong2013,Xu2014}, for which semiconducting two-dimensional (2D) TMDCs have been established as promising candidates. In monolayer TMDCs, the lack of inversion symmetry in 2H-polytypes gives rise to a spin-valley correlation of the band structure which, in combination with strong spin-orbit coupling in those containing heavy transition metals \cite{Zhu2011}, lifts the energy degeneracy of electronic bands of opposite spin polarizations, allowing for valley-selective electronic excitation with circularly polarized light \cite{Mak2012,Zeng2012,Zhu2011,Xiao2012,Cao2012,Berghaeuser2014}. While such an effect should be forbidden in inversion symmetric materials, recent theoretical work suggests that the absence of inversion symmetry within moieties of the unit cell locally lifts the spin degeneracy \cite{Zhang2014,Liu2015PRL}. 
The lack of inversion symmetry and the presence of in-plane dipole moments within individual TMDC layers can be seen as atomic site Dresselhaus and Rashba effects and can cause a hidden spin texture in a globally inversion symmetric material \cite{Zhang2014}. This is supported by the observation of spin-polarized valence bands in 2H-\WSe\ by photoelectron spectroscopy \citep{Yu1999} and spin-resolved ARPES \cite{Riley2014}. 
Polarization-resolved photoluminescence experiments on inversion-symmetric bilayer samples~\cite{Mak2012,Zeng2012,Wu2013,Zhu2014,Jones2014} have shown varying degrees of circular dichroism. This has primarily been explained by symmetry breaking induced by applied or intrinsic electric and magnetic fields.

In this Letter, we demonstrate that in centrosymmetric samples of 2H-\WSe, it is possible to generate spin-, valley- and layer-polarized excited states in the conduction band. By employing time- and angle-resolved photoemission spectroscopy (trARPES) with circularly polarized pump pulses, we observe spin-polarized excited state populations in the K valleys, which is in addition localized to a single \WSe\ layer. TDDFT calculations \cite{Runge1984} confirm the valley- and layer-selectivity of the different pump polarizations and reveal the underlying excited state electronic structure. We identify scattering pathways in momentum space and determine time constants of electronic lifetimes in the valleys of the conduction band. The initial population at the corner (K points) of the Brillouin Zone (BZ) is rapidly transferred to the global minimum of the conduction band at the \Si\ points, where the interlayer coupling is strongly enhanced. 

\begin{figure*}[bth]
\begin{center}
\includegraphics[width=1.0\textwidth]{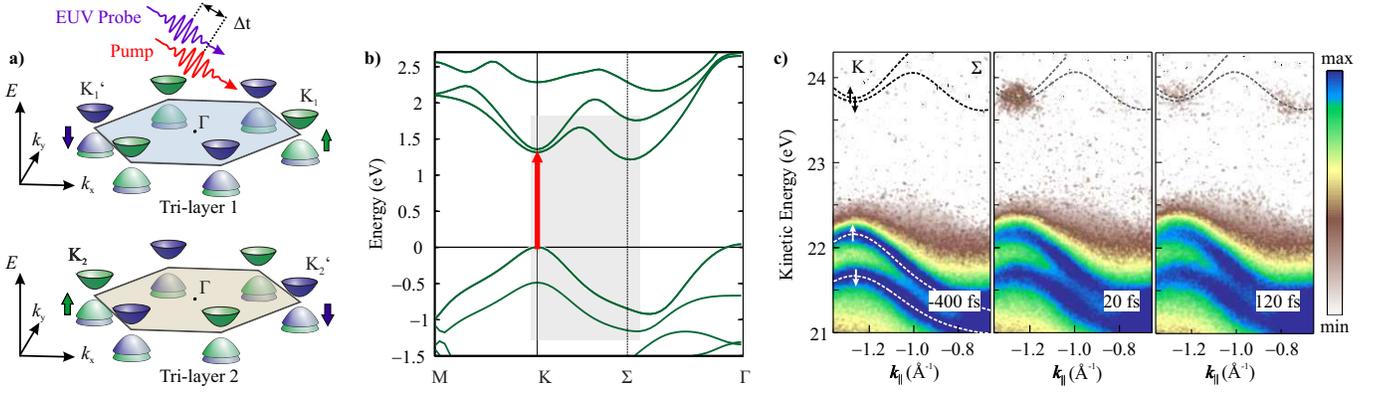}
\caption{{\bf a)} Sketch of the local band structure and the spin-valley correlation at the K and K' points of two non-interacting layers with 2H stacking. The two layers together comprise the unit cell of both bilayer and bulk crystal, in which the spin-polarization vanishes for every $\mathbf{k}$-point. {\bf b)} Electronic band structures of bilayer \WSe\ calculated by DFT. The red arrow and the shaded rectangle indicate the optical excitation and the experimentally investigated part of the electronic structure, respectively. {\bf c)} Electron intensity maps at various pump-probe delays obtained with linearly polarized pump pulses (logarithmic colour scale). The overlaid valence and conduction bands are obtained by DFT calculated for a bilayer, with the conduction bands shifted 250~meV upwards to match the experimental observations.}
\label{fig:estr}
\end{center}
\end{figure*}

The electronic bands and the corresponding spin polarizations at the K points of two decoupled layers of \WSe\ with 2H stacking are depicted in Figure \ref{fig:estr}a. As the valence states in the K valleys are almost 100\% spin-polarized and well localized within the atomic layers \cite{Riley2014,Finteis1997}, it is possible to relate spin-polarized valleys to individual layers \cite{Jones2014}. The integrated spin-polarization, however, vanishes for every k-point due to the inversion symmetry of the unit cell. The band structure of an inversion-symmetric bilayer of \WSe, as calculated by DFT \cite{SI, Gonze2009, Perdew1981, Andrade2015, Hartwigsen1998}, is shown in Figure \ref{fig:estr}b. The global minimum of the conduction band \Si\ is located about halfway between \Ga\ and K, in agreement with previous work \cite{Finteis1997,Riley2015}. The conduction band states are shifted in energy by 250~meV to match the experimental data. 
We employ trARPES with femtosecond extreme ultraviolet (EUV) probe pulses of 23~eV photon energy \cite{Rettenberger1997, SI}
in order to probe the response of the electronic structure to optical excitation along the \Si-K and \Si'-K' lines in the BZ within an energy window including the spin-split valence bands and the lowest conduction bands. 
\replaced{While both pump and probe pulses penetrate many layers,}{With this photon energy,} the mean free path of photo-emitted electrons \added{with the given energy} is shorter than one tri-layer, resulting in a very high surface sensitivity \cite{Riley2014}. 
This allows experimental access to the electron dynamics in the top layer, i.e.~the upper half of the topmost unit cell of the sample, as illustrated in Figure \ref{fig:estr}a. 
In combination with spin-selective excitation, our experimental approach  provides valley- and layer-sensitive information of the excited state population. 
The pump pulses are tuned into resonance with the A exciton absorption at 1.63~eV \cite{Frindt1963}, corresponding to direct optical transitions at the K points from the upper valence band to the conduction band, as indicated by the red arrow in Figure~\ref{fig:estr}b. The splitting of the valence bands of 500~meV ensures valley-sensitivity of the excitation. The applied fluences are on the order of few mJ/cm$^2$ and result in excitation densities above the threshold for the excitonic Mott transition \cite{Chernikov2015}, therefore creating an electron-hole plasma. Three representative photoelectron intensity maps obtained using linearly polarized pump-pulses are shown in Figure \ref{fig:estr}c. The kinetic energy scale has been corrected for the analyser work function and the maps  have been overlaid with the calculated band structure of the bilayer, indicating the position of the conduction band valleys at K and \Si, which are initially unoccupied. Upon optical excitation, states at the K points are resonantly excited (middle panel) and this population is transferred to the \Si\ valleys at longer delay times, as shown in the third panel. 

\begin{figure*}[bth]
\begin{center}
\includegraphics[width=1.0\textwidth]{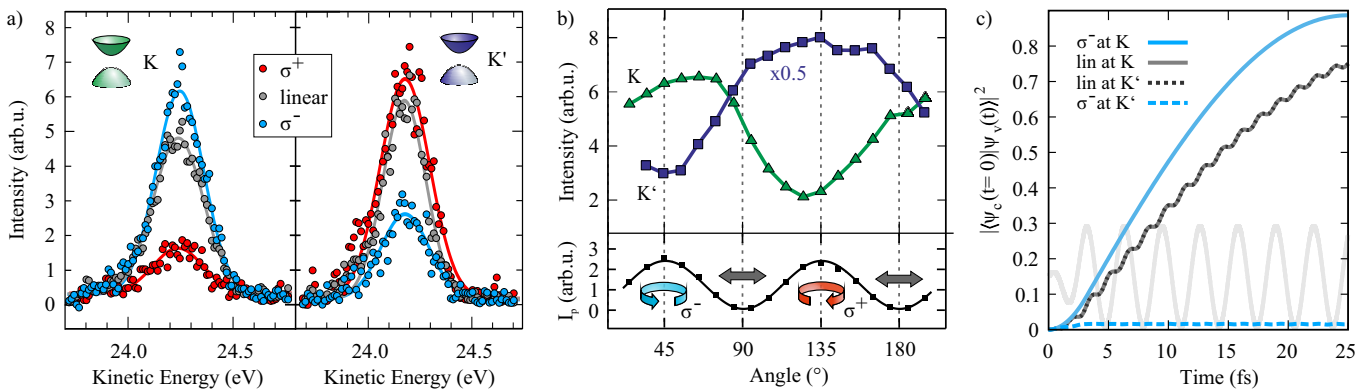}
\caption{{\bf a)} Energy distribution curves of the excited state signal at K and K', 15~fs after excitation with linearly and circularly polarized optical pump pulses. 
{\bf b)} Excited state photoelectron intensity at the K point (green triangles) and the K' point (blue squares) at a fixed pump-probe delay as a function of the angle of a quarter-wave plate in the pump beam. 
The lower panel shows the measurement (squares) and fit (lines) of the parallel component of the pump polarization, from which the polarization state in dependence of the wave plate angle is deduced. 
{\bf c)} Evolution of excited state population at the K and K' points in the upper half of a \WSe\ bilayer for linear and circular ($\sigma^-$) pump polarization from TDDFT simulations. The light grey oscillation indicates the applied electric field of the simulation.}
\label{fig:dichroism}
\end{center}
\end{figure*}

We now monitor the excited state population in the K valleys after excitation with circularly polarized light. The energy distribution curves (EDCs) of the conduction band at the K point, taken at a delay of $\Delta$t=15~fs, are presented in Figure \ref{fig:dichroism}a for excitation with light of both circular helicities and of linear polarization. 
A strong circular dichroic effect is evident in the excited state population with $\sigma^-$ light exciting significantly more electrons in the K valley compared to $\sigma^+$, and excitation with linear polarization being intermediate. The excited state population in the non-equivalent K' valley shows the opposite effect with respect to helicity.
The difference in the observed contrast$| \frac{\sigma^{+}-\sigma^{-}}{\sigma^{+}+\sigma^{-}}| $ at K and K' is likely caused by two factors related to the rotation of the sample between measuring at both points in $k$-space: at K', the polarization state of the pump pulses deviate from purely circular due to non-normal incidence; in addition, the probed area is larger, making contributions from surface areas with opposite spin texture to the integrated signal more likely. 
The continuous evolution of excited state populations in the K and K' valleys in dependence of the polarization state is shown in Figure~\ref{fig:dichroism}b, showing an oscillatory dependence of the excited state signal on the angle of a quarter-wave plate, which gradually changes the polarization of the pump pulses from circular left to linear p-polarized to circular right as indicated by the arrows. 
The minima and maxima of the excited state population occur at angles corresponding to circular polarizations, and maximal excitation at the two inequivalent K points occurs for opposite helicity. 
Figure~\ref{fig:dichroism}c shows the simulated evolution of the valence electronic structure projected on the ground state conduction band of the upper layer of bilayer \WSe\ under different pump polarizations \cite{SI}. Whereas linear polarized light results in a build-up of excited state population at both K and K' points, we observe a pronounced dichroism of the layer-projected excited state population at the corners of the BZ for circular polarization. We confirmed that this population build-up is also visible in an ab initio simulation of the pump-probe ARPES process. The experiments together with the TDDFT calculations unambiguously demonstrate the ability to valley- and spin-selectively excite electrons in bulk 2H-\WSe, additionally leading to a layer-pseudospin polarization.

\begin{figure}[bth]
\begin{center}
\includegraphics[width=1.0\columnwidth]{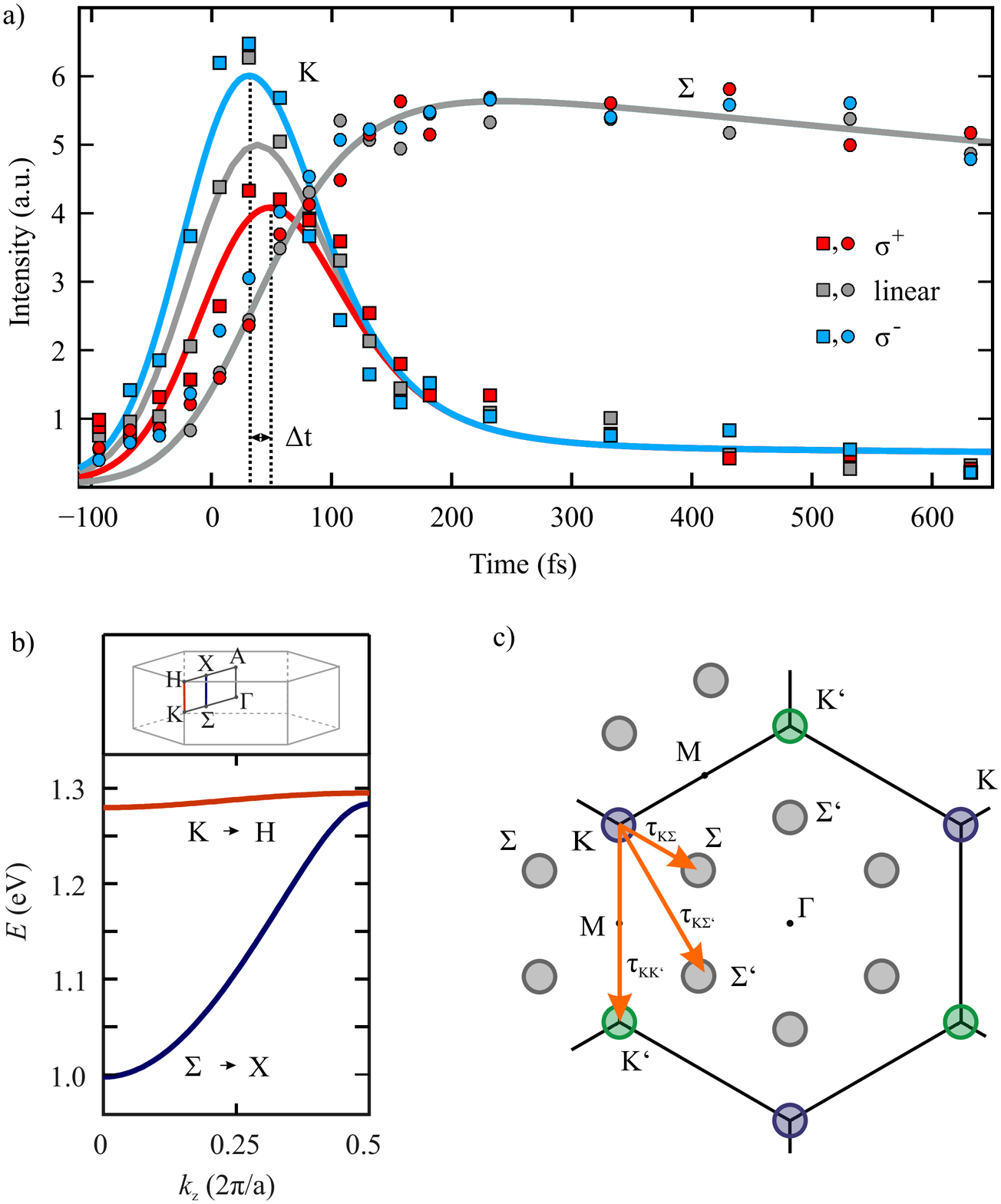}
\caption{{\bf a)} Temporal evolution of the excited state signal in the K (squares) and \Si\ (circles) valleys for different pump polarizations. The solid lines are taken from a scattering model (see text).  {\bf b)} Top: sketch of the BZ of bulk \WSe. Bottom: Dispersion of the conduction band states in bulk \WSe\ along the K-H and \Si-X directions as defined above.  {\bf c)} Cut through the \Ga-K-M plane of the Brillouin zone of a single \WSe{} tri-layer including K and \Si\ valleys. The orange arrows indicate possible electronic scattering processes between the valleys.}
\label{fig:tev}
\end{center}
\end{figure}

The measured dichroic contrast at the K points (Figure \ref{fig:dichroism}) critically depends on pump-probe delay and diminishes at long delay times. The temporal evolution of the photoelectron intensity in the conduction band is shown in Figure \ref{fig:tev}a for $\sigma^+$, linear and $\sigma^-$ pump polarizations. 
In the K valleys, the dichroism is observed through the different build-up of electronic population. It is pronounced at early delays and is subsequently lost as electrons scatter away from the K points. 
The build-up and maximum of electronic population are slightly delayed for the circular pump polarization which predominately excites the opposite (K') valleys (dashed vertical lines in Figure~\ref{fig:tev}a). This suggests that electronic scattering between the non-equivalent K valleys is significant and contributes to the rapid reduction of dichroic signal with delay. 
The decay of population at the K points is accompanied by an increase of signal at the \Si\ points, as carriers scatter towards the global minimum of the conduction band. In monolayers, these states are energetically higher, and directly couple to the substrate \cite{Cabo2015}. 
The dynamics of the populations in the \Si\ valleys are, within our experimental accuracy, independent of pump helicity, despite the dichroic population in the K valleys. This is explained by the different orbital character of states at the \Si\ valleys, which our DFT calculations predict to be significantly more delocalized between the layers as revealed by the 20-fold increase of the $k_z$-dispersions shown in Figure~\ref{fig:tev}b \cite{SI}.

To investigate the scattering pathways and quantify the electronic dynamics, the time dependence of the electronic populations are modelled with a set of rate equations, assuming an initial valley-selective excitation of carriers \cite{SI}. 
Figure \ref{fig:tev}c shows the scattering processes included in the model. 
As states at \Si\ are delocalized along the z-direction, they equally fill by scattering from K in one layer and K' in the neighbouring layer, making these indistinguishable. 
By numerically optimizing the parameters to the six measured time-traces (Figure \ref{fig:tev}a), the dynamics are well reproduced (solid lines) for $\tau_{K\Sigma}$ =$\tau_{K\Sigma'}=(70\pm15)$ fs and $\tau_{KK'} = (60\pm30)$ fs. 
The \Si\ valley populations are long-lived and remain observable up to several tens of picoseconds. 
Our observations visualize efficient intervalley scattering as the main reason for the strongly quenched luminescence \cite{Mak2010} of multilayer TMDC samples and contradict the assumption of stable excitons in bulk \WSe\ \cite{Langer2016}. The measured intervalley timescales are compatible with carrier-carrier \cite{Schmidt2016} as well as electron-phonon scattering \cite{Steinhoff2016}. While the identification of the dominant mechanism requires complementary experiments such as time-resolved diffraction, these scattering processes have been suggested to be spin-conserving \cite{Liu2015,Liu2015PRL}. We therefore expect a high degree of spin polarization in the conduction band even after intervalley scattering.


\begin{figure}[bth]
\begin{center}
\includegraphics[width=1.0\columnwidth]{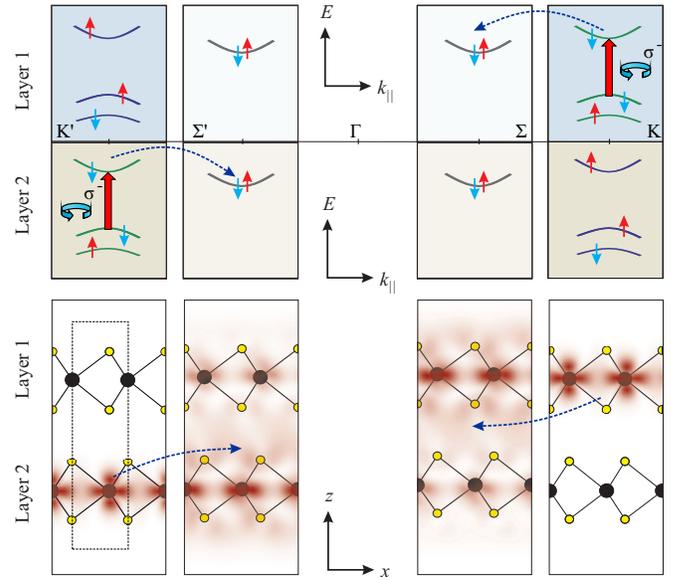}
\caption{Sketch of the electronic processes in reciprocal space (top panel) and real space electron densities integrated over one in-plane direction of the wave functions at the respective $\mathbf{k}$-states in the conduction band (bottom panel). The large red arrows symbolize the electronic excitation as a consequence of $\sigma^-$-polarized pump pulses. Electronic scattering from the K to the \Si\ valleys is indicated by the dashed blue arrows. The spin character of the bands is represented by the small blue and red arrows. The dashed box in the lower left panel marks the real space unit cell.}
\label{fig:conc}
\end{center}
\end{figure}
The key findings of our work are summarized in Figure~\ref{fig:conc} by  real- and reciprocal-space illustrations of the involved electronic states. The lower panels show the excited state charge densities in the K and \Si\ valleys of the conduction band obtained by integration over one in-plane direction of the wave functions~\cite{SI}.
Optical excitation with circularly polarized pump pulses generates spin-polarized excited state population with transient valley and layer polarization, i.e.~localization in reciprocal space to specific K valleys (upper panels) and in real space to individual trilayers (lower panels). 
This effect is a consequence of the atomic site asymmetries~\cite{Zhang2014,Riley2014,Liu2015PRL} in TMDCs, which is predicted to occur in a range of materials including topological insulators and superconductors \cite{Zhang2014}. Intervalley scattering on the sub-100~fs time scale populates states in the \Si\ valleys with pronounced three-dimensional character, leading to a loss of valley and layer polarization. In view of utilizing multilayer \WSe~as a source of ultrafast spin currents, efficient electronic coupling of the acceptor states to the conduction band states at \Si\ is required.
This suggests a strategy for the extraction of spin-polarized carriers between neighbouring layers in TMDC multilayers and heterostructures \cite{Geim2013, Withers2015, Ye2016} where transfer of electrons between layers is governed by the state-dependent interfacial electronic coupling, which can be controlled by an appropriate choice of materials, stacking order, and relative orientation. Such control combined with microscopic understanding of electron dynamics, as provided here, are crucial for conceiving TMDC-based spintronic device concepts.

Recently, two studies reporting ultrafast electron dynamics in the related compound MoS$_2$ were published \cite{Hein2016, Wallauer2016}.

This work was funded by the Max Planck Society, the European Research Council (ERC-2010-AdG-267374), by the Ministerio de Econom\'ia y Competitividad (FIS2013-
46159-C3-1-P), and Grupos Consolidados UPV/EHU
(IT578-13). Access to the Artemis Facility was funded by Laserlab-Europe III (EU-FP7, grant agreement No. 284464). R.B. thanks the Alexander von Humboldt Foundation for financial support. H.H. acknowledges support from the People Programme (Marie Curie Actions) of the European Union's Seventh Framework Programme FP7-PEOPLE-2013-IEF project No. 622934. C.M. acknowledges support by the Swiss National Science Foundation under Grant No. PZ00P2\_154867.


%

\end{document}